\documentclass[12pt,letterpaper]{article}
\usepackage[sort&compress, numbers]{natbib}
\usepackage{graphicx}
\usepackage{amssymb}
\usepackage[margin=1in,top=0.5in,bottom=0.6in]{geometry}
\usepackage{lineno}
\usepackage{enumitem}
\usepackage{aas_macros}

\usepackage{hyperref}
\usepackage{sidecap}
\usepackage{cite}

\begin{document}

\begin{titlepage}
	\centering
	{\large\bfseries Astro2020 Science White Paper\par}
	\vspace{0.6cm}
	{\scshape\LARGE Physics Beyond the Standard Model With Pulsar Timing Arrays \par}
	\vspace{0.5cm}
	\includegraphics[width=0.55\textwidth]{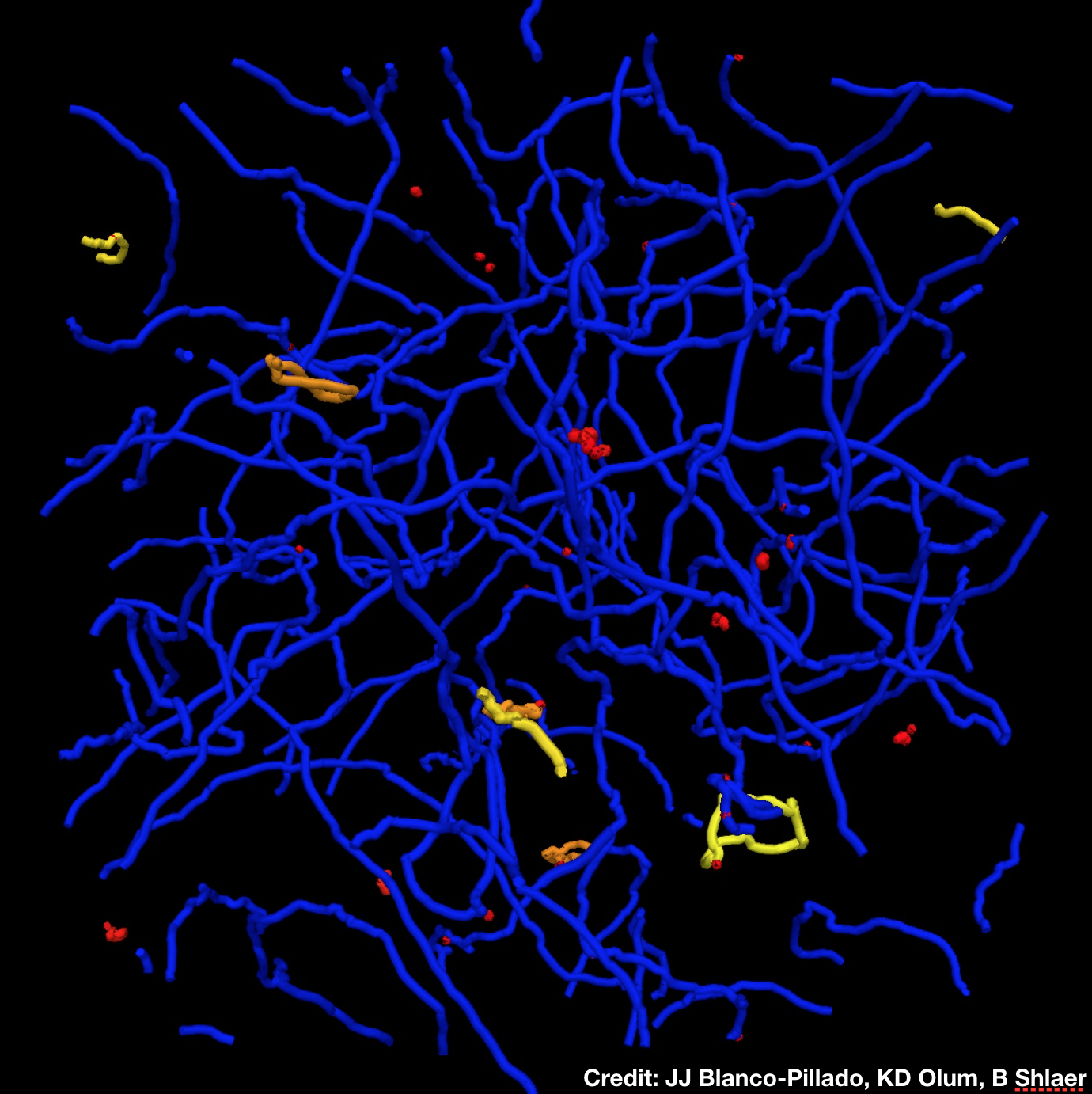}\par\vspace{0.5cm}
	\textbf{Principal authors}: Xavier Siemens \textit{(University of Wisconsin -- Milwaukee)}, {\normalsize\href{mailto:siemens@uwm.edu}{siemens@uwm.edu}}, Jeffrey~Hazboun \textit{(University of Washington -- Bothell)}, {\normalsize\href{mailto:hazboun@uw.edu}{hazboun@uw.edu}}
	\par
	\vspace{0.3cm}

		    \textbf{Co-authors:} Paul~T.~Baker \textit{(West Virginia University)},
		    Sarah~Burke-Spolaor \textit{(West Virginia University/Center for Gravitational Waves and Cosmology/CIFAR Azrieli Global Scholar)},
		    Dustin~R.~Madison \textit{(West Virginia University)},
		    Chiara~Mingarelli \textit{(Flatiron Institute)},
		    Joseph Simon \textit{(Jet Propulsion Laboratory, California Institute of Technology)},
		    Tristan~Smith \textit{(Swarthmore College)}.

	\vspace{0.3cm}
	{\normalsize This is one of five core white papers written by members of the NANOGrav Collaboration.} \vspace{-0.5cm}
	\paragraph{\textbf{Related white papers}}
	\begin{itemize}[noitemsep,topsep=0pt]
	    \item \textit{Nanohertz Gravitational Waves, Extreme Astrophysics, And Fundamental Physics With Pulsar Timing Arrays}, J.~Cordes, M.~McLaughlin
	    \item \textit{Supermassive Black-hole Demographics \& Environments With Pulsar Timing Arrays}, S.~R.~Taylor, S.~Burke-Spolaor
	    \item \textit{Multi-messenger Astrophysics with Pulsar Timing Arrays}, L.Z.~Kelley, M.~Charisi et al.
	    \item \textit{Fundamental Physics With Radio Millisecond Pulsars}, E.~Fonseca, P.~B.~Demorest, S.~M.~Ransom
	\end{itemize}
	
	\vspace{0.3cm}
	
	\noindent \textbf{Thematic Areas:} \hspace*{60pt} 
	$\square$ Planetary Systems \hspace*{10pt} 
	$\square$ Star and Planet Formation \hspace*{10pt}\linebreak
    $\square$ Formation and Evolution of Compact Objects \hspace*{10pt} 
    ${\rlap{$\checkmark$}}\square$ Cosmology and Fundamental Physics %\linebreak
    $\square$  Stars and Stellar Evolution \hspace*{1pt} $\square$ Resolved Stellar Populations and their Environments \hspace*{40pt} %\linebreak
     $\square$    Galaxy Evolution   \hspace*{45pt} 
     $\square$  Multi-Messenger Astronomy and Astrophysics %\hspace*{65pt} %\linebreak
\end{titlepage}

\subsection*{Executive summary}

Pulsar timing arrays (PTAs) will enable the detection of nanohertz gravitational waves (GWs) from a population of supermassive binary black holes (SMBBHs) in the next $\sim 3-7$ years. In addition, PTAs provide a rare opportunity to probe exotic physics. Potential sources of GWs in the nanohertz band include 
\begin{itemize}
    \item {\bf cosmic strings and cosmic superstrings},
    \item {\bf inflation}, and,
    \item {\bf phase transitions in the early universe}.
\end{itemize}
GW observations will also make possible
{\bf tests of gravitational theories}
that, by modifying Einstein's theory of general relativity, attempt to explain the origin of cosmic acceleration and reconcile quantum mechanics and gravity, two of the most profound challenges facing fundamental physics today. Finally, PTAs also provide a new means to probe certain
{\bf dark matter} 
models. 

\vspace{0.3cm}
\noindent
Clearly, {\bf a positive detection of any of these  observational signatures would have profound consequences for cosmology and fundamental physics}. In this white paper we describe these potential signatures.  

\vspace{0.5cm}

\begin{SCfigure}[][h]
\centering
\vspace{-0.15cm}
\caption{Plot of cosmic (super)string GW spectra for values of the dimensionless string tension  $G\mu/c^2$ in the range of $10^{-23}$-$10^{-9}$, as well as the spectrum produced by SMBBHs, along with current and future experimental constraints. PTA sensitivity will not be superseded until the LISA mission scheduled for launch in 2034. The Big Bang Observer (BBO) is a future planned space-based GW detector. Figure from Ref.  ~\citep{Blanco-Pillado:2017rnf}.}
\includegraphics[width=0.6\textwidth]{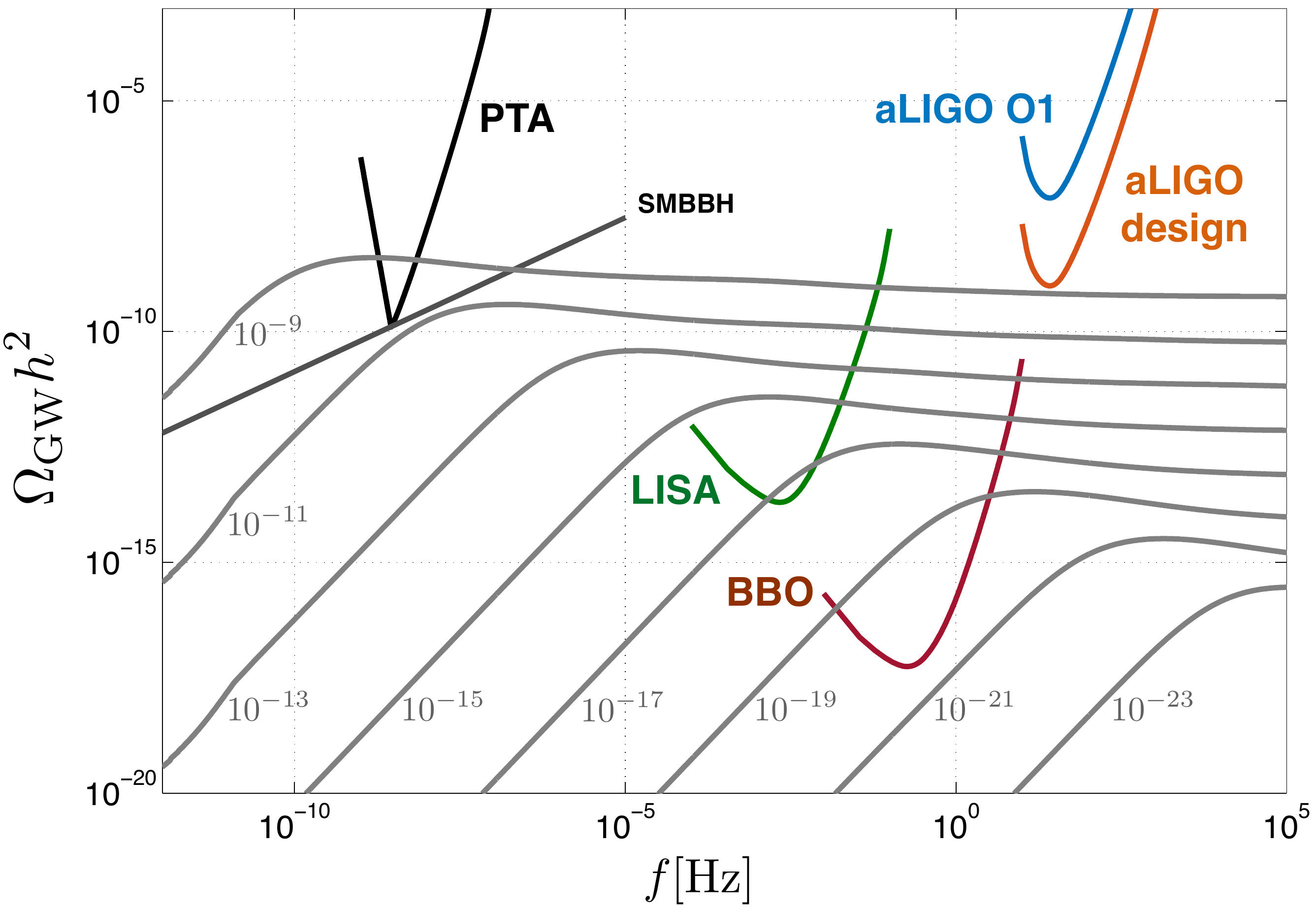}
\label{fig:omega}
\end{SCfigure}

\subsection*{Cosmic strings and cosmic superstrings}
\label{sec:strings}

Cosmic strings are topological defects that can form during phase transitions in the early Universe~\citep{kibble,alexbook}, and cosmic superstrings are the fundamental strings of string theory stretched to cosmological scales due to the expansion of the Universe~\citep{jones1,tye,dvaliandvilenkin,jones2,copeland,jackson}. In a cosmological setting, and for the most simple superstring models, both cosmic string and superstring networks evolve in the same way. For a detailed review of cosmic (super)string network evolution and observational signatures see \citep{Copeland:2009ga} and references therein. Cosmic (super)strings can exchange partners when they meet and produce loops when they self-intersect. These loops then oscillate and lose energy to GWs generating bursts and a stochastic background~\citep{gammaraybursts,DV0,DV1,DV2,burstprd,stochprl}---the signals we wish to detect~\citep{maggiore-pr,Blanco-Pillado:2017rnf}. Strings are characterized by their mass per unit length $\mu$, which is normally given in terms of the dimensionless parameter $G\mu/c^2$, the ratio of the string energy scale to the Planck scale squared. {\bf The detection of a stochastic background from cosmic (super)strings, or GWs from individual cosmic (super)string loops, would be transformative for fundamental physics}.

\vspace{0.3cm}
\noindent
 {\bf PTAs are currently the most sensitive experiment for the detection of cosmic (super)strings and will remain so for at least the next decade and a half}. Correspondingly, 
pulsar-timing experiments are producing the most constraining bounds on the energy scale and other model parameters of cosmic strings and superstrings. As of the writing of this white paper, the best limit on the string tension, $G\mu/c^2 < 5.3(2) \times 10^{-11}$, is several orders of magnitude better than constraints from cosmic microwave background (CMB) data, and comes from the NANOGrav Collaboration~\citep{2018ApJ...859...47A}. Figure \ref{fig:omega} shows the stochastic background spectrum produced by cosmic strings in terms of the dimensionless density parameter $\Omega$ versus frequency for dimensionless string tensions $G\mu/c^2$ in the range $10^{-23}$-$10^{-9}$. Overlaid are current and future experimental constraints from PTAs, ground-based GW detectors, and spaced-based detectors. PTA sensitivity will not be superseded until the LISA mission which is scheduled for launch in 2034.

\subsection*{Primordial gravitational waves from inflation}
\label{sec:inflation}

The evolution of the very early Universe is thought to include a period of exponential expansion called
inflation that accounts for the observed homogeneity,  isotropy, and flatness of the Universe~\citep{Brout:1977ix,Starobinsky:1980te,Kazanas:1980tx,Sato:1980yn,Guth:1980zm,Linde:1981mu,Albrecht:1982wi}. Additionally, by expanding quantum fluctuations present in the pre-inflationary epoch, inflation seeds the density fluctuations that evolve into the large scale structures we see in the Universe today~\citep{Mukhanov:1981xt,Hawking:1982cz,Guth:1982ec,Starobinsky:1982ee,Bardeen:1983qw}, and produces a stochastic background of GWs~\citep{Starobinsky:1979ty,Rubakov:1982df,Abbott:1984fp}. This background of GWs is broad-band, like the one produced by cosmic strings, and potentially detectable by multiple experiments. 

\vspace{0.3cm}
\noindent
Detecting primordial GWs from inflation has been a critical objective of CMB experiments for some time, see~\citep{2016ARA&A..54..227K} and references therein. The CMB is sensitive to the lowest frequency portion of the GW spectrum from inflation, and CMB data can be used to constrain the tensor-to-scalar ratio, which is the ratio of the size of GWs produced to that of scalar perturbations (which seed density fluctuations as described above).  For standard inflation models the GW background in the PTA band is likely to be fainter than that of SMBBHs, though that depends in part on the character of the SMBBH spectrum at the lowest frequencies where environmental effects like accretion  from a circumbinary disk or stellar scattering can reduce SMBBH GW emission~\citep{Sampson:2015ada}.
In addition, some inflationary models have a spectrum that rises with frequency.  Thus, GW detectors operating at higher frequencies than CMB experiments, like {\bf PTAs and space- and ground-based interferometers, can be used to constrain the shape of the inflationary GW spectrum}. Indeed, PTA, CMB, and GW interferometer data across 29 decades in frequency have already begun to place stringent limits on such models~\citep{2016PhRvX...6a1035L}.

\subsection*{Phase transitions in the early universe}
\label{sec:PTs}

The early Universe may have experienced multiple phase transitions as it expanded and cooled. Depending on the detailed physical processes that occur during a phase transition, GWs can be generated with wavelengths of order the Hubble length at the time of the phase transition. That length scale, suitably redshifted, translates into a GW frequency today. Thus, GW experiments at different frequencies today probe horizon-sized physical processes occurring at different times in the early Universe, with higher frequency experiments probing earlier and earlier times.

\begin{SCfigure}
\centering
\vspace{-0.5cm}
\caption{Dashed lines show the GW spectrum of a first order QCD phase transition for various phase transition durations, along with PTA sensitivities. LISA will
not be able to detect a signal from the QCD phase transition; the electroweak phase transition that occurs at higher temperatures, earlier times, and therefore maps onto higher frequencies, is a more promising source for LISA. Figure from Ref.  ~\citep{2010PhRvD..82f3511C}.}
\includegraphics[width=0.6\textwidth]{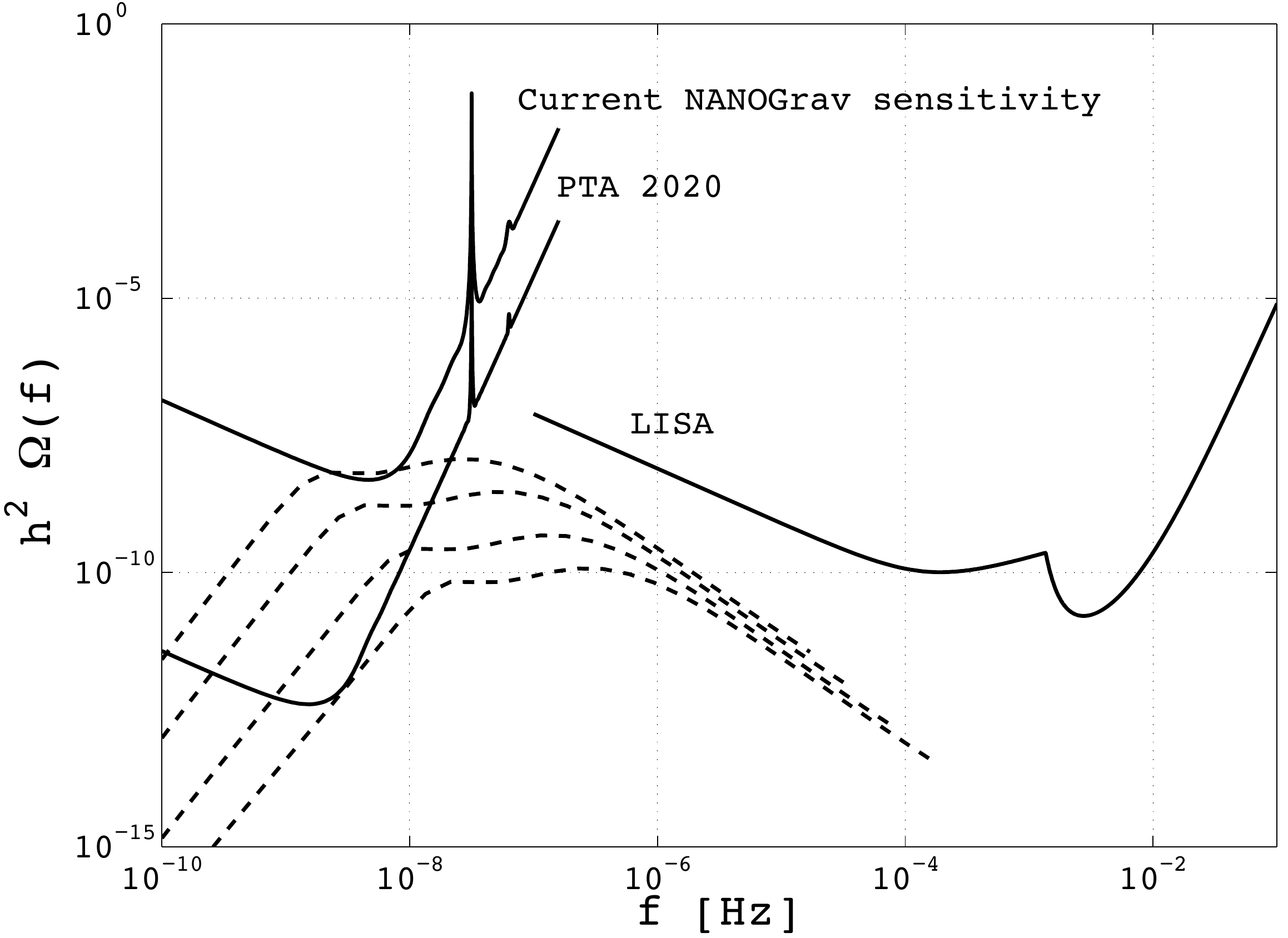}
\label{fig:qcd}
\end{SCfigure}

\vspace{0.3cm}
\noindent
{\bf The nanohertz frequency band accessible to PTAs maps onto the era in the early universe when the quantum chromodynamics (QCD) phase transition took place}, about $10^{-5}$~s after the Big Bang. The horizon at that time was on the order of 10~km, and any GWs generated at that length scale at that time would today be stretched to about 1~pc (or 3 light-years), which corresponds to GW frequencies of about 10~nHz, and lie within the PTA sensitivity band.  The possibility that interesting QCD physics can result in a GW signal detectable by PTAs was first pointed out by Witten in the 1980s~\citep{PhysRevD.30.272}.
More recently Caprini et al.~\citep{2010PhRvD..82f3511C} considered the possibility of a first order phase transition at the QCD scale. In standard cosmology the
QCD phase transition is only a cross-over, and we do not expect it to generate GWs. However, if the neutrino chemical potential is sufficiently large it can become first order (it is worth pointing out that if sterile neutrinos form the dark matter, we expect a large neutrino chemical potential). Thus PTAs provide a window onto physical processes occurring in the universe at the time of the QCD phase transition, and could detect GWs from a first order phase transition at that time (see Fig.~\ref{fig:qcd}). As can be seen in Fig.~\ref{fig:qcd} a given PTA will become sensitive to lower frequencies as the baselines on their data sets increases through the next decade.

It should be noted that all of these stochastic background signals (cosmic strings, inflation, and phase transitions) will have to contend with the background of super-massive black hole binaries expected to be detected in the next few years by PTAs. However, {\bf long enough baselines and  sufficiently distinguishable spectral characteristics will make these signatures individually resolvable} \citep{Parida:2015fma}. 

\subsection*{Gravitational-wave tests of General Relativity}
\label{sec:altgrav}

Attempts to explain the origins of cosmic acceleration and to reconcile gravity and quantum mechanics, two outstanding fundamental physics problems, often involve modifications to Einstein's theory of general relativity. Testing general relativity as a theory of gravity is therefore a crucial goal for PTAs~\citep{Yunes:2013dva}. Here we focus on tests of general relativity made possible by PTA detections of gravitational waves; strong-field tests of GR based on 
binary neutron star orbits are also possible with PTA data (see white paper by Fonseca, et al., {\it Fundamental Physics with Radio Millisecond Pulsars})

\begin{SCfigure}
\centering
\caption{The six possible GW polarizations in metric theories of gravity. The solid and dotted lines in each case represent the effect of the GW on a freely falling ring of masses at integer and half-integer multiples of the GW period. General relativity predicts only plus and cross modes (shown on the left in red), every other metric theory of gravity predicts the existence of more polarizations.  Finding evidence in favor of scalar or vector polarizations would immediately rule out general relativity. Reproduced from \citep{Chamberlin:2011ev}}
\includegraphics[width=0.6\textwidth]{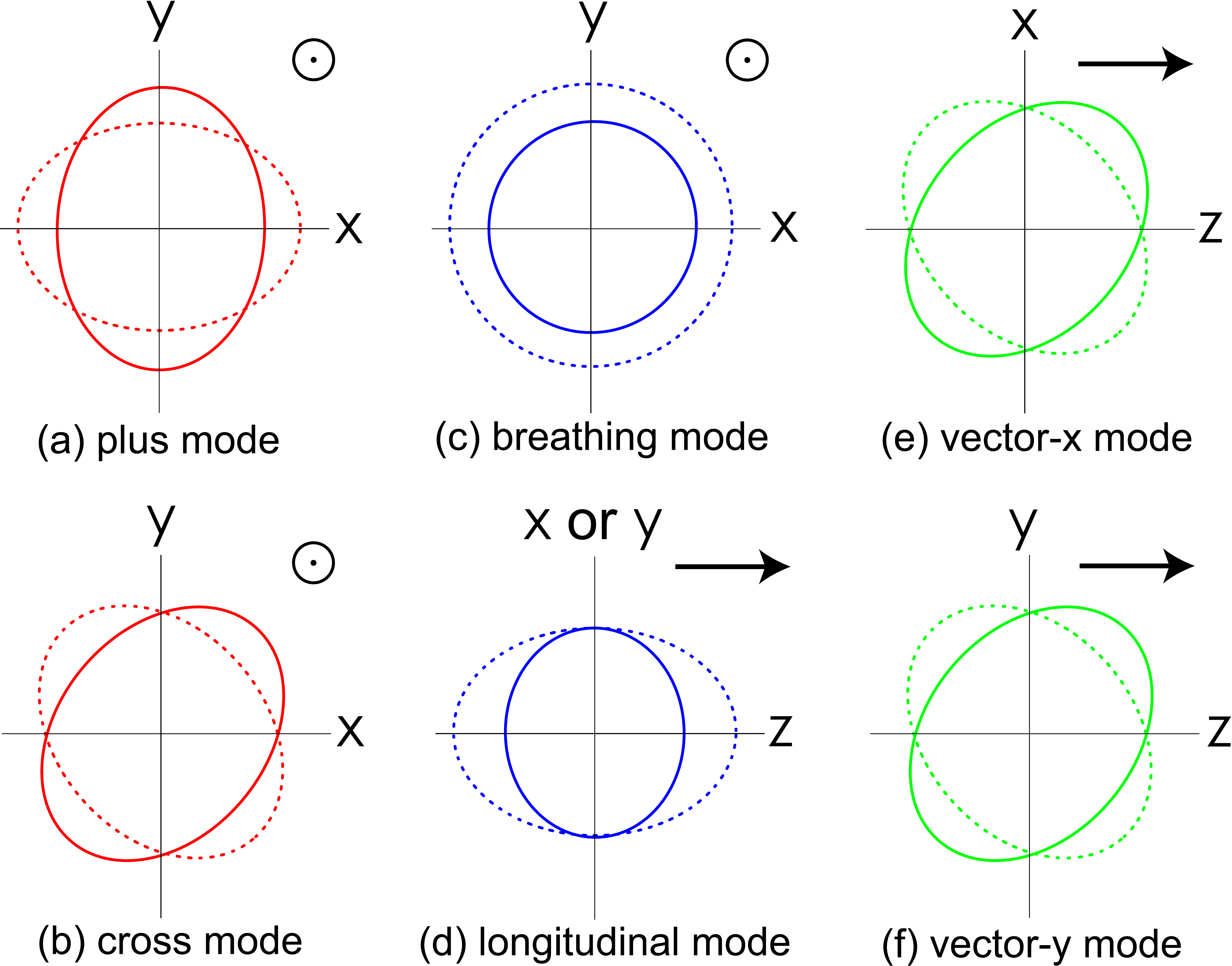}
\label{fig:pols}
\end{SCfigure}
\vspace{0.3cm}
\noindent
General relativity predicts the existence of GWs which travel at the speed of light, are transverse, and have two polarizations.  Other metric theories of gravity generically predict the existence of GWs with different properties: additional polarizations and modified dispersion relations. Metric theories of gravity can have up to six possible GW polarizations~\citep{Eardley:1973br}. These six components are classified according to their transformation properties under rotations, resulting in two scalar polarizations, two vector polarizations, and two tensor polarizations \citep{Eardley:1973br,Eardley:1974nw} (see Fig.~\ref{fig:pols} for an illustration). The two tensor components, the only ones predicted to exist by general relativity, are commonly referred to as the plus and cross polarizations. {\bf Finding experimental evidence in favor of non-tensor polarizations and/or non-standard dispersion properties for GWs would immediately rule out general relativity}. 

\vspace{0.3cm}
\noindent
PTAs offer significant advantages over interferometers like LIGO for detecting new polarizations or constraining the polarization content of GWs. Each line of sight to a pulsar can be used to construct an independent projection of the various GW polarizations, and
since PTAs typically observe tens of pulsars, linear combinations of the data can be formed to measure or constrain each of the six polarizations many times over~\citep{Yunes:2013dva}. Additionally, PTAs have an  enhanced response to the longitudinal polarization~\citep{Chamberlin:2011ev}. Indeed, the constraint on the energy density of longitudinal modes inferred from PTA data is about three orders of magnitude better than the constraint for the transverse modes~\citep{Cornish:2017oic}. Using the NANOGrav 9-year data set~\citep{Arzoumanian:2015gjs} these authors set the $95\%$ upper limits on the amplitudes of stochastic GW backgrounds from the non-GR polarizations at $\Omega_{TT+ST} h^2 < 7.7\times10^{-10}$,
$\Omega_{VL}h^2 < 3.5\times10^{-11}$ and $\Omega_{SL} h^2 < 3.2\times10^{-13}$, corresponding to the sum of tensor-transverse and scalar-transverse (breathing) modes, the vector-longitudinal modes, the and scalar-longitudinal mode. Note that only the sum of scalar- and tensor-transverse modes can be constrained, see~\citep{Cornish:2017oic}. 

\subsection*{Dark matter}
\label{sec:DM}
Dark matter is an essential component of the universe, accounting for about a quarter of its energy density. Dark matter explains a wide range of cosmological phenomena, from galaxy rotation curves to the detailed characteristics of the CMB and large-scale structure formation. Despite the enormous success of dark matter its nature remains an open question in fundamental physics.

\vspace{0.3cm}
\noindent
{\bf PTA data may help elucidate the constituents and properties of dark matter}. Certain classes of dark matter models produce observable signatures at nanohertz frequencies. 
Scalar fields with masses around $10^{-23}$~eV, for example, can produce periodic oscillations in the gravitational potential with a strain $h \sim 10^{-15}$ and frequencies in the nanohertz range~\citep{kr14}, well within the detectable range of PTAs in the coming decade.  These models are motivated by string theory  axions,  a type of light scalar field. Their small masses result in a macroscopic de Broglie wavelength, which is the origin of the name ``fuzzy'' dark matter. An attractive feature of fuzzy dark matter is that it avoids the cuspy halo problem. PTA data are already being used to constrain such models~\citep{Porayko:2018sfa}. Finally, standard cold dark matter (CDM) models naturally produce small scale clumps which may also be detectable by PTAs. A CDM clump moving near the Earth or a pulsar produces an acceleration that could be measurable in PTA data, providing an opportunity to test the CDM paradigm~\citep{Kashiyama:2018gsh}.

\subsection*{Key Detectors \& Requirements}\label{sec:detectors}

These science opportunities require PTA collaborations such as NANOGrav to monitor $\sim$100 pulsars with weekly to monthly cadences over a wide radio band (several GHz) with large collecting area radio telescopes over a period of time of several decades. NANOGrav has been taking pulsar timing data for a little over  14 years, and currently times 76 millisecond pulsars using about 850 hours per year at the Arecibo Observatory and 550 hours per year at the Green Bank Telescope.

\vspace{0.3cm}
\noindent
Increasing our sensitivity to stochastic backgrounds, such as the background produced by SMBBHs, cosmic strings, or inflation, is best achieved by adding more pulsars to the array, more or less independently of their quality~\citep{Siemens:2013zla}. Increasing our sensitivity to individual GW sources, such as individual black hole binaries and cosmic string loops, and to perform improved dark matter searches and GW tests of General Relativity, requires timing pulsars with the highest possible precision.

\vspace{0.3cm}
\noindent
In the coming decade, these requirements imply the need for larger collecting area ($\gtrsim$ 300-m equivalent diameter) steerable telescopes, operating in the frequency range of a few hundred MHz to a few GHz, and able to dedicate substantial amounts of time ($\sim$ 1000s hours per year) to regular sustained pulsar timing observations. Examples of such instruments include the DSA2000~\citep{DSA2000} and the ngVLA~\citep{ngVLA}. An instrument with these capabilities would allow for dramatic advances in the science opportunities described in this white paper.

\pagebreak
\bibliographystyle{unsrt}
\bibliography{bib.bib}

\end{document}